\documentstyle[aps,pre,epsf,floats]{revtex}
\begin{document}
\draft
\parskip 1mm

\twocolumn[\hsize\textwidth\columnwidth\hsize\csname@twocolumnfalse%
\endcsname

\title  {Critical behavior of roughening transitions in
parity-conserving growth processes}
\author {Haye Hinrichsen$^1$ and G\'eza \'Odor$^2$\\}
\address{$^1$Max-Planck-Institut f\"ur Physik komplexer Systeme,
	 N\"othnitzer Stra{\ss}e 38, D-01187 Dresden, Germany}
\address{$^2$ Research Institute for Technical Physics and
	 Materials Science,
	 P. O. Box 49, H-1525 Budapest, Hungary}
\date   {July 14, 1999}
\maketitle

\begin{abstract}
We investigate a class of parity-conserving
solid-on-solid models which describe the growth
of an interface by the deposition and evaporation of dimers.
As a key feature of the models, evaporation of
dimers takes place only at the edges
of terraces, leading to a roughening transition
between a smooth and a rough phase.
We consider several variants of growth models in
order to identify universal and nonuniversal properties.
Moreover, a parity-conserving polynuclear growth
model is proposed. All variants display the same
type of universal critical behavior at the 
roughening transition.
Because of parity-conservation,
the critical behavior at the first few layers
can be explained in terms of unidirectionally
coupled branching annihilating random walks
with even number of offspring.
\end{abstract}

]


\section{Introduction}

In the present work we continue the investigation of a certain
class of nonequilibrium models for interfacial growth by adsorption
and desorption~\cite{HinrichsenOdor99a}.
The models may be used to describe a layer-by-layer growth process
of a $d$-dimensional surface by deposition and evaporation of
{\it dimers} which are aligned with the surface.
Upon adsorption the dimers dissociate into two atoms at
neighboring lattice sites. It is assumed that the atoms cannot diffuse
on the surface. Furthermore, atoms cannot evaporate
from the interior of completed layers. Only pairs of
neighboring atoms at the {\it edges} of terraces are able to
form a dimer and evaporate back into the gas phase.

The most interesting property of this class of growth processes 
is the emergence of a roughening transition from a smooth to 
a rough phase at a certain critical ratio of the adsorption and 
desorption rates~\cite{Reviews}.
In contrast to equilibrium growth models and nonequilibrium growth
processes described by the Kardar-Parisi-Zhang (KPZ)
equation~\cite{KPZ86}, where roughening transitions take place
only in $d \geq 2$ dimensions, the models discussed in the present work
exhibit a robust roughening transition even in one spatial dimension.
As will be shown below, the scaling properties of the interface at the
transition differ significantly from the usual scaling laws
for roughening interfaces, i.e., the models exhibit anomalous
roughening properties.

The motivation to consider growth processes of dimers 
originates in recent studies of similar interface
models for adsorption and desorption of monomers~\cite{Alon96}.
In these models individual atoms are adsorbed with
probability $p$ on each lattice site, whereas
atoms evaporate with probability $1-p$ solely at the 
edges of terraces. It was shown that monomer models of this kind
exhibit a roughening transition which is
closely related to the universality class of directed
percolation (DP)~\cite{DP}.
A more detailed analysis revealed that the
critical behavior of the first few layers can be
explained in terms of unidirectionally coupled DP
processes~\cite{CoupledDP}. It turned out that such hierarchies of
coupled DP processes not only describe the monomer models
of Ref.~\cite{Alon96} but also various other DP-related
growth processes, including polynuclear growth
(PNG) models~\cite{KW89}, certain models for fungal
growth~\cite{LopezJensen98}, and a recently introduced
model designed for real-space renormalization~\cite{BMGP99}.
Thus, the concept of `coupled DP' characterizes a whole universality
class of roughening transitions~\cite{PB99}.

DP itself is the generic universality class for phase
transitions into trapped (absorbing) states and
is known to be extremely robust with respect to the
choice of the dynamic rules.
The DP conjecture~\cite{JanssenGrassberger}
states that in a random process with short-range interactions 
any transition from a fluctuating active phase into 
a single absorbing state should belong to the
DP class, provided that the dynamics is characterized
by a single-component order parameter without
additional symmetries. Non-DP critical behavior
is expected in systems where one of these requirements
is violated. An  important example is the
so-called parity-conserving (PC) universality class
for phase transitions into absorbing states in which
the parity of the particle number is conserved. In
one dimension, this conservation law can also be interpreted
as a $Z_2$-symmetry between two different absorbing states.
The PC class is represented most prominently by branching
annihilating random walks with two
offspring (BAW2)~\cite{BAWE,BAWMOD,BAWEFT}.
Other examples include nonequilibrium
kinetic Ising models~\cite{Nekim}, interacting monomer-dimer
models~\cite{IMD}, as well as models with two symmetric
absorbing states~\cite{GDK}.
It is therefore near at hand to investigate the question how
the physical properties change if the DP mechanism of 
the growth models in Ref.~\cite{Alon96} is replaced
by a parity-conserving dynamics. To this end we modify
the dynamic rules of these models by using
{\em dimers} instead of monomers which adsorb with probability
$p$ and desorb at the edges of terraces (including
solitary dimers) with probability $1-p$.
As dimers consist of two atoms, the number of particles at
each height level is conserved modulo 2. As shown in
Ref.~\cite{HinrichsenOdor99a}, the model exhibits a
robust roughening transition which is related to the PC class.

In the following we present a detailed numerical analysis of the
critical behavior of parity-conserving growth processes. One of our aims
is to clearly identify universal and nonuniversal properties.
To this end we consider four variants of the model,
a restricted version with random sequential updates, an
unrestricted version as well as the corresponding
variants with parallel updates. It turns out that all variants
exhibit the same type of critical behavior at the roughening transition.
In particular, the critical behavior at low-lying levels can
successfully be described in terms of unidirectionally
coupled PC processes, generalizing the concept of `coupled DP'.
Above the transition, however, restricted and unrestricted variants
display different properties.

The paper is organized as follows. In Sect.~\ref{DefinitionSection}
we define four variants of the dimer growth model. Their
phenomenological properties are described
in Sect.~\ref{PhenoSection}. Sects.~\ref{WidthSection}
and~\ref{FirstFewSect} investigate the critical properties of the
interface width and the densities of exposed sites at the first few
layers. We also discuss aspects of spontaneous symmetry breaking
and unusual scaling properties for random initial conditions.
The critical behavior at the roughening transition can partly be explained 
in terms of unidirectionally coupled PC processes for which
we propose a field-theoretic formulation in Sect.~\ref{CoupledSection}.
As will be shown in Sect.~\ref{PNGSection}, it is even possible to construct
a polynuclear growth model which belongs to the same universality class.
Our conclusions are summarized in Sect.~\ref{ConcSection}.

\section{Models definition}
\label{DefinitionSection}
%
%
\begin{figure}
\epsfxsize=80mm
\centerline{\epsffile{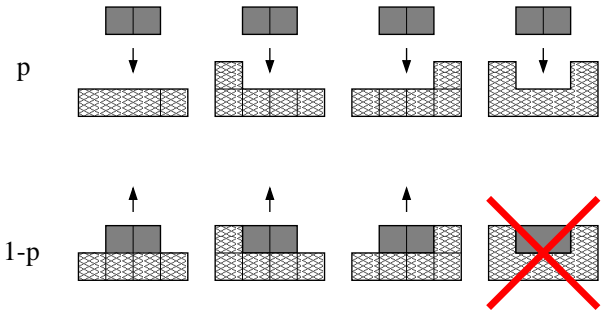}}
\caption{
\label{FigRules}
Variant A in $d=1$ dimension:
Dimers are adsorbed with probability $p$
and desorbed at the edges of terraces
with probability $1-p$. Evaporation from
the middle of plateaus is not allowed.
}
\end{figure}

The class of models may be introduced in terms of a 
$d$-dimensional interface which evolves by adsorption and
desorption of dimers. As a key feature desorption may only take 
place at the edges of a plateau, i.e. at sites which have at
least one neighbor at a lower height.
The dimer growth model is defined on a $d$-dimensional square
lattice with $L^d$ sites and periodic boundary conditions. Each
site $i$ is associated with an integer height variable 
$h_i \in \mbox{\sf Z\mbox{\hspace*{-0.45 em}}Z}$.
We consider four variants A,B,C,D of the model which differ
by their dynamic rules.

Variant A is a restricted solid-on-solid
(RSOS) model evolving by random sequential updates.
For each attempted update a pair of adjacent sites
$i$ and $j$ is selected at random. If the heights
$h_i$ and $h_j$ are equal a dimer is adsorbed with probability $p$
\begin{equation}
\label{Adsorption}
\begin{array}{l}
h_i \rightarrow h_i+1 \\
h_j \rightarrow h_j+1
\end{array}
\end{equation}
or desorbed with probability $1-p$
\begin{equation}
\label{Desorption}
\begin{array}{l}
h_i \rightarrow h_i-1 \\
h_j \rightarrow h_j-1
\end{array}
\qquad
\mbox{if \ }
\min_{k \in <i,j>} h_k < h_i \,,
\end{equation}
where $k$ runs over the nearest neighbors of sites $i$ and~$j$.
An attempted update is rejected if it violated the RSOS constraint
\begin{equation}
\label{Restriction}
|h_i-h_j| \leq 1 \,,
\end{equation}
i.e., the heights at neighboring sites may differ by at most one step.
A Monte-Carlo sweep consisting of $L^d$ local update attempts
corresponds to a time increment $t \rightarrow t+1$.
For $d=1$ the dynamic rules are shown in Fig.~\ref{FigRules}.
Notice that the rules are translationally invariant
in time, space as well as in height direction.
Thus, the layer at $h=0$ has no particular physical meaning, 
although we will often use a flat interface at
zero height as initial condition.

Variant B is an unrestricted solid-on-solid (SOS) model
which is defined by the same rules without
restriction~(\ref{Restriction}). Although unrestricted
growth is less realistic, it exhibits
essentially the same type of critical behavior at the
roughening transition, supporting the claim of universality.

\begin{figure}
\epsfxsize=70mm
\centerline{\epsffile{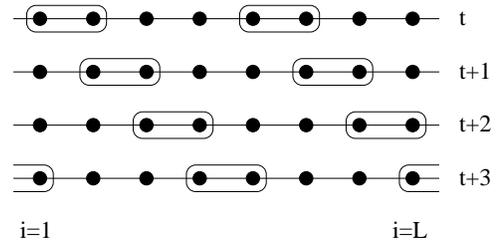}}
\vspace{2mm}
\caption{Variants C and D: Cyclic parallel updates on four different
sublattices in a one spatial dimension.
\label{sublat}
}
\end{figure}
Variants C and D are counterparts of A and~B
employing {\em parallel} updates.
More precisely, their lattice is divided into
several sublattices in a way that synchronous updates
on each sublattice according to the rules~(\ref{Adsorption})
and~(\ref{Desorption}) do not overlap. In $d=1$
dimension at least three sublattices are needed.
However, for technical reasons it is more convenient to 
work with four different sublattices, as illustrated in
Fig.~\ref{sublat}. We implement this update scheme on a
parallel computer with 24000 string processors.
Associating an individual processor with each lattice 
site we are able to perform efficient simulations.
Further details on massive parallel computing are given
in Ref.~\cite{benchmarkpaper}.

\section{Phenomenology of the\\ interface dynamics}
\label{PhenoSection}
%
%
\begin{figure}
\epsfxsize=85mm
\centerline{\epsffile{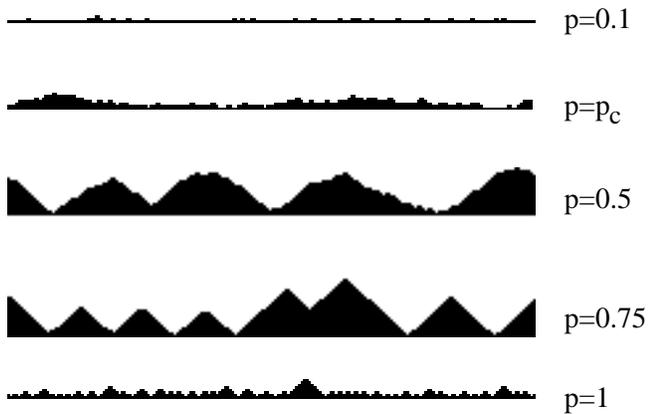}}
\vspace{2mm}
\caption{
Typical interface configurations of the restricted dimer
model (variant A) for various values of $p$
(see text).
\label{FigDemoRest}
}
\end{figure}
Although the dimer models are defined in arbitrary 
spatial dimensions, this work is restricted to the
one-dimensional case $d=1$. Clearly, the morphology
of the interface depends on the growth rate $p$.
Let us first consider the restricted variants
(see Fig.~\ref{FigDemoRest}). If $p$ is very small, only
a few dimers are adsorbed at the surface, staying there 
for a short time before they evaporate back into the
gas phase. Thus, the interface is anchored 
to the actual bottom layer and
does not propagate. As $p$ increases, a growing number
of dimers covers the surface and  large islands of
several layers stacked on top of each other are formed.
Approaching a certain critical threshold $p_c$
the mean size of the islands diverges
and the interface evolves into a rough state.

Above $p_c$ one may expect the interface
to detach from the bottom layer in the same
way as the interface of monomer models starts
to propagate in the supercritical phase.
However, since dimers are adsorbed at neighboring
lattice sites, solitary unoccupied sites may
emerge. These pinning centers prevent
the interface from moving and lead
to the formation of `droplets' (see Fig.~\ref{FigDemoRest}).
Due to interface fluctuations, the pinning centers
can slowly diffuse to the left and to the right.
When two of them meet at the same place, they annihilate
and a larger droplet is formed. Thus, although the
interface remains pinned, its roughness
increases continuously.

As realized in~\cite{nohfac}, a second transition
takes place at $p=0.5$ where the width grows most
rapidly. In this case the droplets
reach an almost triangular shape with unit slope,
i.e., the surface becomes {\em faceted}.
The tilted surface of the droplets
fluctuates predominantly by randomly moving
triplets of sites at equal
height along `staircases' of unit slope.
Inspecting the dynamic rules
it is easy to verify that these `landings'
move upwards with probability $p$
and downwards with probability $1-p$.
This explains why the faceting
transition takes place exactly at $p=0.5$.
For $p>0.5$ the fluctuations are confined
to an exponentially small region at the top
of the droplets. Therefore, the faceted
interface coarsens on a logarithmic time scale.
A pathological situation emerges for $p=1$
where evaporation of dimers is forbidden.
As the pinning centers, once formed, cannot
diffuse the interface quickly evolves
into a frozen configuration.

\begin{figure}
\epsfxsize=85mm
\centerline{\epsffile{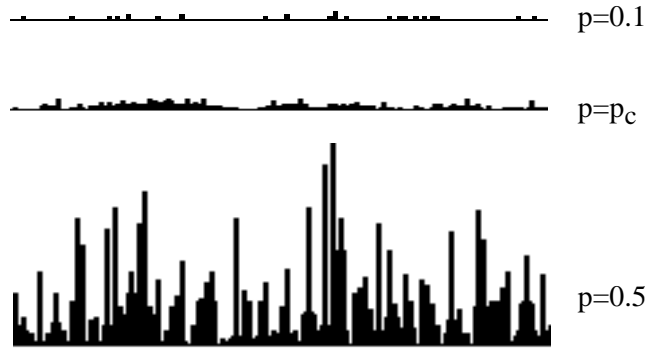}}
\vspace{2mm}
\caption{
Typical interface configurations of the unrestricted dimer
model (variant B) for various values of $p$. At the transition
$p=0.5$ large spikes of stacked dimers are formed. For $p>0.5$ these
spikes are biased to grow with constant velocity.
\label{FigDemoUnrest}
}
\end{figure}

To summarize, the restricted variants A and C
display three different phases, a smooth phase
$p<p_c$, a rough phase $p_c < p < 0.5$, and
a faceted phase $p>0.5$. The phase structure
of the unrestricted variants $B$ and $D$ is very
similar (see Fig.~\ref{FigDemoUnrest}).
They too exhibit a roughening transition
at a certain critical threshold $p_c$. However,
the rough phase and the transition at $p=0.5$
are different in character due to the formation
of spikes, as will be described below.
%
%

\section{Critical behavior of the\\ interface width}
\label{WidthSection}

\noindent
The morphology of a growing interface is usually characterized
by its width
\begin{equation}
W(L,t) =
\Bigl[
\frac{1}{L} \, \sum_i \,h^2_i(t)  -
\Bigl(\frac{1}{L} \, \sum_i \,h_i(t) \Bigr)^2
\Bigr]^{1/2}
\,.
\end{equation}
In order to investigate the scaling properties of the width,
we perform Monte-Carlo simulations starting from a flat
interface $h_i(0)=0$ and get the following results:
\begin{figure}
\epsfxsize=85mm
\centerline{\epsffile{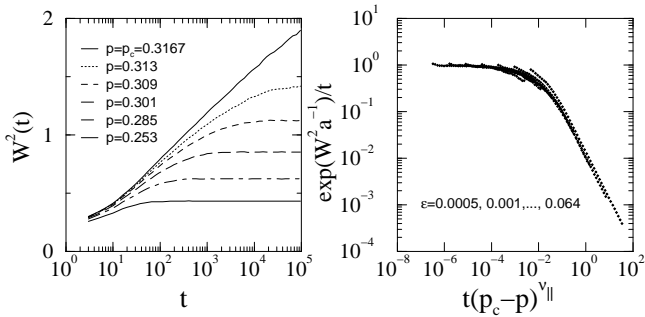}}
\vspace{2mm}
\caption{
Left:
The squared width $W^2(t)$ measured
in variant A for various values of $p\leq p_c$. 
Right: Data collapse according to
Eq.~(\ref{WidthOffScaling}),
visualizing the scaling function~$F$.
Throughout the whole paper $t$ is measured
in units of Monte-Carlo sweeps.
\label{FigOffWidth}
}
\end{figure}

\paragraph{\bf The smooth phase $p<p_c$:}
Fig.~\ref{FigOffWidth} shows the temporal evolution of $W^2(t)$
for different values of $p\leq p_c$
measured in model A with $L=10000$ sites. 
As can be seen, the width first increases as 
$W(t) \sim \sqrt{\ln t}$ until it saturates at some constant value.
The initial logarithmic increase suggests the scaling form
\begin{equation}
\label{WidthOffScaling}
W^2(t,\epsilon) \simeq a \,\ln \Bigl[ t
\, F(t\epsilon^{\nu_\parallel}) \Bigr]\, ,
\end{equation}
where $a$ is an amplitude factor and
$\epsilon=|p-p_c|$ denotes the distance from
criticality.
The exponent $\nu_\parallel$ describes the singular
behavior of the temporal correlation length
$\xi_\parallel \sim \epsilon^{-\nu_\parallel}$ close
to the roughening transition.
$F$ is a universal scaling function
with the asymptotic behavior
\begin{equation}
\label{AsymtoticBehavior}
F(\zeta) = \left\{
\begin{array}{ll}
const \, \, & \mbox{if} \,\,\, \zeta \rightarrow 0 \\
\zeta^{-1} &  \mbox{if} \,\,\, \zeta \rightarrow \infty
\end{array}
\right. \, .
\end{equation}
In order to estimate $a$ and $\nu_\parallel$ we
adjust these quantities in a way that the curves for
$\exp(W^2/a)/t$ versus $t\epsilon^{\nu_\parallel}$
collapse onto a single one.
Using the estimates $p_c=0.317(1)$,
$\nu_\parallel = 2.4(4)$, and $a=0.17(1)$
we obtain a fairly convincing data collapse
(see right hand graph of Fig.~\ref{FigOffWidth}).
However, we observe considerable deviations for larger
values of $\epsilon$. These deviations may
indicate corrections of the scaling
form~(\ref{WidthOffScaling}) in the off-critical
regime and will be analyzed in Sect.~\ref{OffCritSim}.
Similar results are obtained for variants B,C, and D.

\paragraph{\bf The roughening transition at $p=p_c$: }

\begin{figure}
\epsfxsize=85mm
\centerline{\epsffile{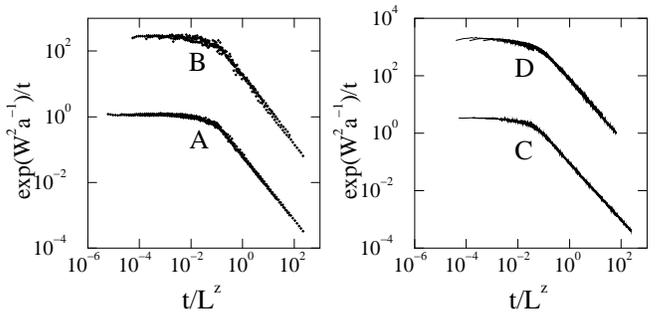}}
\caption{
\label{FigFSWidth}
Finite-size scaling of the interface
width $W(L,t)$ for variants A-D. The
graphs show data collapses according to the scaling
form~(\ref{WidthFSScaling}), visualizing
the scaling function~$G$.
}
\end{figure}

At the roughening transition the interface width of an infinite
system grows {\em logarithmically} as $W \sim \sqrt{\ln t}$.
A similar logarithmic behavior has also been observed at the
roughening transition of monomer models~\cite{Alon96}.
The logarithmic time dependence suggests the finite-size scaling
form~\cite{Comment1}
\begin{equation}
\label{WidthFSScaling}
W^2(L,t) \simeq a \, \ln \Bigl[ t \, G(t/L^z) \Bigr]\, ,
\end{equation}
where $z$ denotes the dynamic exponent.
$G$ is a universal scaling function
with the same asymptotic behavior as in
Eq.~(\ref{AsymtoticBehavior}).
Thus, the interface of finite systems is expected to
saturate at a constant value $W(L) \sim \sqrt{\ln L}$.
Our numerical results are summarized in Fig.~\ref{FigFSWidth}.
Plotting $\exp(W^2/a)/t$ against $t/L^z$ in
the time interval $10 \leq t \leq 10^5$ for 
system sizes $L=32,64,\ldots,4096$, we determine
$a$ and $z$ by data collapse. For all variants we obtain
accurate data collapses. The estimates for $z$ coincide
in all cases, indicating universal properties of the
roughening transition independent of the RSOS
constraint and the type of updates (see Table~\ref{Table}).
In fact, as we will argue in the following Section,
$z\simeq 1.75$ is the dynamic exponent of the PC
universality class. The amplitude $a$, however, is nonuniversal.
A very accurate data collapse is obtained by
simulations of variant C on a parallel computer
and highly supports the validity of the scaling
form~(\ref{WidthFSScaling}).
It would be interesting to confirm this scaling form
by direct diagonalization of the transfer matrix~\cite{Mendonca}.

\paragraph{\bf The rough phase $p_c < p < 0.5$:}

In the rough phase the formation of pinning
centers prevents the interface from propagating.
By diffusion and pairwise annihilation
of the pinning centers the width increases
very slowly. Since the width displays a rather inconclusive
scaling behavior, it was impossible
conjecture an appropriate scaling form~\cite{Comment2}.
It seems that the width initially increases
algebraically until it slowly crosses over to a
logarithmic increase $W(t) \sim \sqrt{a \ln t}$.
The amplitude $a$ and the crossover time grow with $p$
and diverge at the transition at $p=0.5$.
The crossover time provides a typical
time scale of the dynamics in the rough phase.
Apart from that the roughening transition
is associated with another temporal correlation length
$\xi_\parallel=(p-p_c)^{\nu_\parallel}$.
Thus, the rough phase is characterized by a
complicated interplay of at least two
different time scales.

\paragraph{\bf The transition at $p=0.5$:}
%
%
\begin{figure}
\epsfxsize=80mm
\centerline{\epsffile{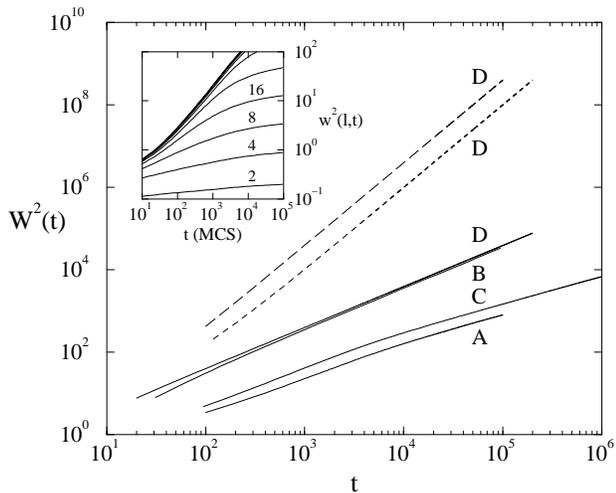}}
\vspace{2mm}
\caption{
Power-law roughening of the width in large systems.
The figure shows the squared width as a
function of time at the transition $p=0.5$ (solid lines),
as well as above the transition for $p=0.6$ (dashed line) and
$p=0.7$ (long-dashed line). The inset shows the squared local width
of variant A for increasing box sizes
$l=2,4,8,16,\ldots$ as a function of time (see text).
\label{powrough}
}
\end{figure}
The restricted as well as the unrestricted variants
undergo a second phase transition at $p=0.5$
where the width increases {\em algebraically} with time as
$W \sim t^{\tilde{\beta}}$. Let us first consider the 
restricted case where the transition can be interpreted
as a {\em faceting transition} between a rough and
a faceted phase~\cite{nohfac}. As the width increases
algebraically, we expect ordinary {\em Family-Vicsek}
scaling~\cite{FamilyVicsek85} of the form
\begin{equation}
W(L,t) \sim L^{\tilde{\alpha}} f(t/L^{\tilde{\alpha}/\tilde{\beta}})\, ,
\end{equation}
where $\tilde{\alpha}$ and $\tilde{\beta}$
are the roughening and the growth exponents, respectively.
However, finite size simulations
(not shown here) reveal a more complex behavior. 
After an initial short-time regime we observe a 
long transient extending over two decades in time
where the interface roughens with the exponents 
$\tilde{\alpha}\simeq 0.5$ and  $\tilde{\beta} \simeq 0.45$.
After approximately $5000$ time steps both variants A and~C
cross over to a different regime with $\tilde{\beta}\simeq 0.33$.
By a finite size data collapse we find a roughening exponent
$\tilde{\alpha}\simeq 1.2$ and a large dynamic exponent
$\tilde{z}= \tilde{\alpha}/\tilde{\beta} \simeq 3$. These
results are consistent with the findings of Ref.~\cite{nohfac}.
In Fig.~\ref{powrough} the crossover appears
as a slight change of the slope, suggesting the faceting
transition at $p=0.5$ not to be fully scale-invariant
but characterized by a finite time scale.
Moreover, as shown in the inset of
Fig.~\ref{powrough}, the faceting transition displays
anomalous scaling properties~\cite{LRC97}, i.e., the
{\em local} width $w(l,t)$ measured in boxes of sizes
$l=2,4,8,16,\ldots$ does not saturate after a short
time. Even for small box sizes it continues to increase
over several time decades until it saturates due
to the RSOS constraint.

Let us now consider the unrestricted variants B and D.
As can be seen in Fig.~\ref{FigDemoUnrest}, their
phase transition at $p=0.5$ is different in character.
Since large spikes are formed, the surface roughens much
faster with a growth exponent of $\tilde{\beta}\simeq 0.5$
(see Fig.~\ref{powrough}). 
In fact, the interface evolves into configurations with
large columns of dimers separated by pinning centers.
These spikes can grow or shrink almost independently.
Thus, the interface roughens by a purely diffusive
mechanism as $W(t) \sim \sqrt{t}$. As the columns are
spatially decoupled, the width does not saturate
in finite systems, i.e., the dynamic
exponents $\tilde{\alpha}$ and $\tilde{z}$ have
no physical meaning.

\paragraph{\bf The faceted/free phase:}

For $p>0.5$ the restricted models A and C evolve into faceted 
configurations (see Fig.~\ref{FigDemoRest}). As shown
in Ref.~\cite{nohfac}, the width first increases
algebraically until the pinning centers become
relevant and the system crosses over to a logarithmic
increase of the width. Therefore, the faceted phase may be
considered as a rough phase. The unrestricted models B and D,
however, evolve into spiky interface configurations. The
spikes are separated and grow independently by deposition of
dimers. Therefore, the interface width increases {\em linearly}
with time, defining the {\em free} phase of the unrestricted models.
\begin{table}
\begin{center}
\begin{tabular}{||c||c|c|c|c||}
variant		& A & B & C & D \\ \hline

restriction 	& yes & no & yes & no  \\
updates         & random & random & parallel & parallel \\ \hline
$p_c$  		& $0.3167(2)$ & $0.292(1)$ & $0.3407(1)$ & $0.302(1)$ \\
$a$	  	& $0.172(5)$ & $0.23(1)$ & $0.162(4)$ & $0.19(1)$ \\
$z$		& $1.75(5)$ & $1.75(5)$ & $1.74(3)$ & $1.77(5)$ \\
$\delta_0$	& $0.28(2)$ & $0.29(2)$ & $0.275(10)$ & $0.29(2)$ \\
$\delta_1$	& $0.22(2)$ & $0.21(2)$ & $0.205(15)$ & $0.21(2)$ \\
$\delta_2$	& $0.14(2)$ & $0.14(3)$ & $0.13(2)$ & $0.14(2)$ \\
$\theta/\nu_\parallel$	& $0.765(10)$ & $0.765(10)$ & $0.753(3)$ & no result \\
\hline
$\tilde{\alpha}$& $1.2(1)$ & undefined & 1.25(5) & undefined \\
$\tilde{\beta}$	& $0.34(1)$ & $0.50(1)$ & $0.330(5)$ & $0.49(1)$ 

\end{tabular}
\end{center}
\caption{
\label{Table}
Numerical estimates for the four variants of the dimer model
at the roughening transition $p=p_c$ (upper part) and the
transition $p=0.5$ (lower part).
}
\end{table}
%
%

\section{Critical properties of the\\ first few layers}
\label{FirstFewSect}
\subsection{Relation to the PC class}

In this Section we investigate the roughening transition at $p=p_c$
in more detail. In order to understand its relation to the
PC class let us consider all sites at the bottom layer
$h_i=0$ as $A$-particles. Adsorption and desorption processes
correspond to certain effective reactions of the $A$-particles. 
For example, the adsorption of a dimer
at the bottom layer corresponds to a  pair-annihilation 
process $2A\rightarrow \O$ at rate~$p$.
Similarly, when a dimer evaporates, two $A$-particles
are created. However, since dimers can only evaporate at the
edges of terraces, this process always requires
the presence of another neighboring $A$-particle, 
giving rise to an effective reaction
$A\rightarrow 3A$ at rate $1-p$. These two processes compete
one another and resemble a branching-annihilating random
walk with two offspring (BAW2)~\cite{BAWE,BAWMOD}
which is known to belong to the PC universality class. 
In one spatial dimension the PC 
class is characterized by three critical exponents
\begin{equation}
\beta=0.92(2), \qquad \nu_\parallel=3.22(6), \qquad \nu_\perp=1.83(3)\,.
\end{equation}
The two scaling exponents $\nu_\parallel$ and $\nu_\perp$ are associated
with temporal and spatial correlation lengths
\begin{equation}
\xi_\parallel \sim |p_c-p|^{-\nu_\parallel} \, , \qquad
\xi_\perp \sim |p_c-p|^{-\nu_\perp} \,
\end{equation}
which diverge at the transition.
The order parameter of the PC transition is the density
of $A$-particles, corresponding to the density $n_0$ of exposed
sites at the bottom layer. Thus, in the smooth phase 
$n_0$ should scale as
\begin{equation}
n_0 \sim (p_c-p)^{\beta_0} \,,
\end{equation}
where $\beta_0=\beta$ is the density exponent of the PC class.
More generally, we expect $n_0$ to obey the scaling form
\begin{equation}
n_0(t,L,\epsilon) \sim t^{-\beta_0/\nu_\parallel} \,
\Phi_0\Bigl(t\, \epsilon^{\nu_\parallel}, \ t/L^z\Bigr)
\ ,
\end{equation}
where $\Phi_0$ is a universal scaling function. As will be shown below,
numerical simulations confirm the validity of this scaling form.
This may be surprising since the $A$-particles do not evolve
independently, but are coupled to dynamic processes
at higher levels of the interface. However, in the present models
this feedback from higher levels is comparatively weak and does
not seem to affect the critical behavior of the $A$-particles at the
roughening transition.

In order to investigate the critical behavior at higher levels,
let us introduce the densities of sites $i$ with $h_i \leq k$
\begin{equation}
n_k = \frac{1}{L} \sum_{j=0}^{k} \sum_i \, \delta_{h_i,j}
\,. \qquad
k=1,2,\ldots
\end{equation}
Following the ideas of Ref.~\cite{Alon96} we expect $n_1,n_2,\ldots$
to scale in the same way as $n_0$ with certain critical exponents
$\beta_k$, $\nu_{\perp,k}$, and $\nu_{\parallel,k}$.
However, since the correlation lengths of neighboring layers
should coincide, the exponents $\nu_{\perp,k}$ and
$\nu_{\parallel,k}$ should not depend on $k$, leading
to the scaling form
\begin{equation}
\label{ScalingForm}
n_k(t,L,\epsilon) \sim t^{-\beta_k/\nu_\parallel} \,
\Phi_k\Bigl(t\, \epsilon^{\nu_\parallel}, \ t/L^z\Bigr)
\ .
\end{equation}
Our numerical results support the validity of this
scaling form for $k\geq 1$ in an intermediate
scaling regime.

\subsection{Estimation of $p_c$ and $\delta_k$}
%
%
\begin{figure}
\epsfxsize=85mm
\centerline{\epsffile{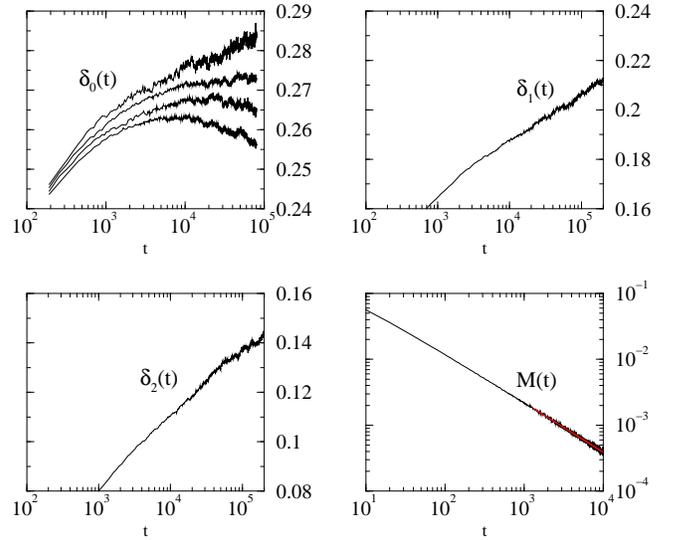}}
\vspace{2mm}
\caption{
\label{aspbpn}
Effective exponents $\delta_k(t)$ measured in model C with $24000$ sites
averaged over $10^4$ samples. The first panel shows $\delta_0(t)$
with $p$ ranging from $0.3406$ to $0.3409$ from top to bottom,
leading to the estimates $p_c=0.3407(1)$ and $\delta_0=0.275(10)$.
The curves for $\delta_1(t)$ and $\delta_2(t)$ at $p=p_c$ are
shown in panel 2 and 3. The last panel shows the magnetization $M$
as a function of time.
}
\end{figure}
\noindent
In order to determine the critical exponents accurately,
we perform time-dependent simulations of variant C on
a parallel computer. Using a large lattice of $24000$ sites
(associating one processor with each lattice site)
we measure $n_0,\ldots,n_2$ as functions of time up to $2\cdot 10^5$
time steps averaged over $10^4$ samples.
At criticality, the densities are expected to decay as
\begin{equation}
n_k(t) \sim t^{-\delta_k}\,,
\end{equation}
where $\delta_k=\beta_k/\nu_\parallel$. In Fig.~\ref{aspbpn}
the effective exponent $\delta_0(t)$ (i.e., the local slope
of $n_0(t)$ in a log-log plot, see~\cite{GT79})
is plotted against $t$ in a logarithmic scale
for various values of $p$. As can be seen,
the curve that appears to become approximately horizontal
for large $t$ corresponds to the critical point
$p_c=0.3407(1)$. For the exponent $\delta_0$ we obtain
the estimate
\begin{equation}
\label{Delta0}
\delta_0=0.275(10)\, ,
\end{equation}
which is in agreement with the expected value
of the PC class $\delta=\beta/\nu_\parallel=0.285(5)$.

The estimation of the exponents $\delta_k$ at higher levels is more
difficult. Averaging $\delta_k(t)$ over two time decades
we obtain the estimates
\begin{equation}
\label{Deltak}
\delta_1=0.205(15)
\, , \qquad
\delta_2=0.13(2) \, .
\end{equation}
Less accurate but compatible results for the other variants
are listed in Table~\ref{Table}. Thus, the
critical behavior at the first few layers is universal
and does not depend on the RSOS constraint and the type of updates.
However, as can be seen in Fig.~\ref{aspbpn}, the effective exponents
$\delta_1(t)$ and $\delta_2(t)$ do not saturate at a constant value
in the long-time limit. Instead they continue to increase
on a logarithmic time scale. At present it is not clear whether
the effective exponents will eventually saturate after very long time.
The drift of the exponents may indicate violations
of scaling in the long-time limit which are 
neither related to finite-size effects nor
to numerical errors in the value of $p_c$. 
Thus, the estimates~(\ref{Deltak}) have to be taken with care.
Similar violations of scaling have been observed in the monomer case
and seem to be an intrinsic property of roughening transitions driven
by absorbing-state transitions. We will come back to this problem in
Sect.~\ref{CoupledSection}.

\subsection{Spontaneous symmetry breaking}

The growth processes considered in this paper are translationally
invariant in height direction. This symmetry is
spontaneously broken in the smooth phase where the models
select one of the heights as the bottom layer of the interface.
In order to quantify this symmetry breaking we consider an
order parameter
\begin{equation}
\label{Magnetization}
M = \frac{1}{L} \sum_j (-1)^{h_j} \,,
\end{equation}
where $j$ runs over all lattice sites. This order parameter 
can take positive or negative values and thus 
reminds of a magnetization. In the stationary
state of the smooth phase $|M|$ is positive and vanishes
at the roughening transition. As in the monomer case,
we expect the magnetization to scale near the
roughening transition as
\begin{equation}
|M| \sim \epsilon^\theta \,,
\end{equation}
where $\theta$ is a critical exponent.
Thus, starting from a flat interface, $|M|$
should decrease at criticality as
$t^{-\theta/\nu_\parallel}$. Performing numerical simulations
(see Fig.~\ref{aspbpn} and Table~\ref{Table}) we find
$\theta/\nu_\parallel \simeq 0.76$ for both the restricted and the
unrestricted variants, suggesting that $\theta$ is a universal
critical exponent. As will be shown in the following
Section, this exponent can also be seen
in unidirectionally coupled PC processes.
Its numerical value $\theta\simeq 2.5$ is much larger than in the
monomer case where the value was found to be $0.65$~\cite{Alon96}.
At present it is not known whether $\theta$ is
independent or related to the other bulk exponents
$\beta,\nu_\parallel,\nu_\perp$.

%
%
\subsection{Finite-size simulations at criticality}
%
%
\begin{figure}
\epsfxsize=75mm
\centerline{\epsffile{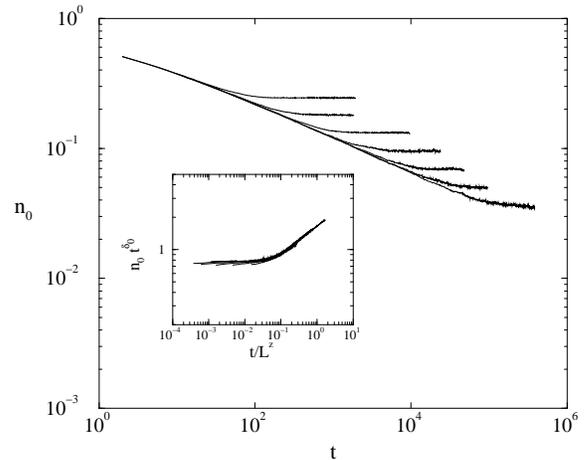}}
\vspace{2mm}
\caption{
\label{fsspar}
Finite size results for the parallel model for $L=64, 128 ...4096$.
The density at the bottom layer $n_0(t)$ is plotted as a function
of time. The inset shows a data collapse which is explained in the text.
}
\end{figure}
The dynamic exponent $z=\nu_{\parallel}/\nu_\perp$ 
can easily be verified by finite-size simulations.
According to the scaling form~(\ref{ScalingForm}),
several measurements of $n_0(t)$ for different system sizes
should collapse onto a single curve
if $n_0 \, t^{\delta_0}$ is plotted versus $t/L^z$.
As shown in Fig.~\ref{fsspar}, the best data collapse is
obtained for $\delta_0\simeq 0.27$ and $z\simeq 1.7$,
which is in agreement with the known PC exponents
$\delta=0.285(5)$, $z=1.76(3)$,
and the previous estimate of $z$ in Sect.~\ref{WidthSection}.

\subsection{Off-critical simulations}
\label{OffCritSim}

In order to determine the exponent $\beta_0$
directly, we perform off-critical simulations in the steady
state of the smooth phase. Here we expect the stationary
density $n_0$ to scale as $(p_c-p)^{\beta_0}$,
where $\beta_0 \simeq 0.92(2)$ is the order parameter exponent of
the PC class. Surprisingly, a rough estimate of $\beta_0$ in
standard Monte-Carlo simulations of variants A and C yields
much smaller values of about $0.6$
which deviate from the expected value by more than
$30\%$. Similar deviations are observed if
$\nu_\perp$ and $\nu_\parallel$
are determined in off-critical simulations. For example, the
estimate $\nu_\parallel=2.4(5)$ in Sect.~\ref{WidthSection}
is much smaller than the expected PC value
$\nu_\parallel\simeq 3.2$.

\begin{figure}
\epsfxsize=85mm
\centerline{\epsffile{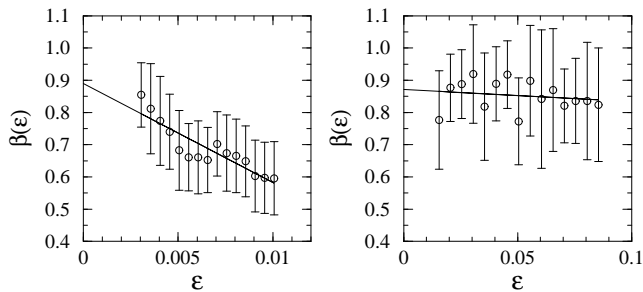}}
\vspace{2mm}
\caption{
\label{drift}
Left: The effective exponent $\beta_0(\epsilon)$ as a function
of~$\epsilon$ in model C with $L=24000$ sites.
After an equilibration of $4\times10^{5}$ time steps
the bottom layer density is averaged over $10^3$ samples.
A simple linear fit gives the estimate $\beta_0 \approx 0.89$.
Right: Corresponding simulation results for the truncated model.
Here the effective exponent is almost independent of $\epsilon$.
Notice the different ranges of $\epsilon$.
}
\end{figure}

Off-critical steady state simulations are known to be
quite inaccurate because of long transients. Moreover,
they are extremely sensitive to errors in the estimation
of $p_c$. However, in the present case the deviations
have a different origin. This can be shown by determining
the effective exponent $\beta_0(\epsilon)$
which is defined as the local slope of
$n_0(\epsilon)$ in a log-log representation between
the data points $(i-1,i)$
\begin{equation}
\beta_0 (\epsilon_i) = \frac {\ln n_{0,i} -\ln n_{0,i-1}}
	      {\ln \epsilon_i - \ln \epsilon_{i-1}} \ \ ,
\end{equation}
where $\epsilon_i = p_c - p_i$, providing an estimate
\begin{equation}
\beta_0 = \lim_{\epsilon\to 0} \beta_0(\epsilon) \,.
\end{equation}
As shown in Fig. \ref{drift}, the effective exponent
{\em increases} slowly and tends towards
the expected PC value as $\epsilon \to 0$.
A simple linear extrapolation yields the estimate $0.9$
which is in rough agreement with the expected value
of~$\beta_0$.

The drift of $\beta_0(\epsilon)$ is unusual and may
be related to the influence of dynamic processes
at higher levels of the interface.
To support this hypothesis, we analyze the
{\em truncated} version of model C where
the heights are restricted to take the values $0$ and $1$
in order to eliminate the feedback from higher levels.
Repeating the same type of analysis at the critical
point $p_c=0.583(1)$ we observe $\beta_0(\epsilon)$
to be in agreement with the expected PC value over a wide
range of $\epsilon$. Thus we are led to the
conclusion that dynamic processes at higher
levels are responsible for corrections to
scaling in the off-critical regime.
%
%
%
\subsection{Unusual scaling for random initial conditions}
%
%
So far we have considered the temporal evolution starting
from an entirely flat interface. It would be natural to expect
the asymptotic scaling behavior to be universal for
any initial interface configuration with finite width and 
short-range correlations. Surprisingly this is not true. 
For example, starting with random initial conditions
$h_i=0,1$ the densities $n_k$ turn out to decay much slower.
For restricted variants we observe an algebraic decay of $n_0$
with an exponent $\delta_0 \simeq 0.13$ which
differs significantly from the value $0.275$ for flat initial
conditions. Similarly, the critical properties of the
faceting transition at $p=0.5$ are affected by random initial
conditions.

The nonuniversal behavior for random initial conditions is
related to an {\em additional} parity conservation law. 
In fact, the dynamic rules (\ref{Adsorption})-(\ref{Desorption}) not only
conserve parity of the particle number but also conserve parity of
the droplet size. Starting with a flat interface the lateral size of
droplets is always even, allowing them to evaporate entirely.
However, for a random  initial configuration,  droplets of odd size
may be formed which have to recombine in pairs before they can evaporate,
slowing down the dynamics of the system.

In the language of BAW2's the additional parity conservation law
is due to the absence of nearest-neighbor diffusion.
Particles can only move by a combination of offspring production
and annihilation, i.e., by steps of {\em two} lattice sites. Therefore,
particles at even and odd lattice sites have to be distinguished.
Only particles of different parity can annihilate. Starting with
a fully occupied lattice all particles have alternating parity
throughout the whole temporal evolution, leading to the usual
critical behavior at the PC transition. For random initial conditions,
however, particles of equal parity cannot annihilate, slowing
down the decay of the particle density. BAW2's and other particle 
processes with parity-conserving diffusion will be analyzed
in a future study.

%
\subsection{Comparison with other models} 
%
%
Recently, Park and Khang proposed a growth model which involves 
two symmetric particle species~\cite{ParkKahng}. Particles
are adsorbed (desorbed) with probability $q$ ($1-q$)
except at those sites where the nearest neighbors are
occupied by particles of the other (same) species. Thus,
the dynamic rules mimic a PC transition with two
symmetric absorbing states~\cite{GDK}.
However, the critical exponents at the roughening transition were
found to differ significantly from the expected PC values.
In fact, once a new layer is completed, the configuration
of the previous layer becomes frozen. Thus the kinks
between different domains do not act as pinning centers
which destroys the PC transition. Consequently,
the interface starts to propagate for $q>q_c$.

Another parity-conserving model introduced by Noh et.~al.~\cite{noh}
was inspired by interacting monomer-dimer models~\cite{IMD}
exhibiting a PC transition. The dynamic rules involve diffusion,
order-preserving branching, and order-breaking branching.
In absence of order-breaking branching, the interface
evolves within a monolayer and undergoes a {\em pre-roughening}
transition belonging to the PC class.
In presence of order-breaking processes several
layers may be formed. However, in this case the interface is
always rough and the transition is lost.

Even more recently, the same authors investigated a class
of parity-conserving growth models~\cite{nohfac}, generalizing
the rules (\ref{Adsorption})-(\ref{Desorption})
by the addition of a process for the
evaporation of dimers from the middle of plateaus.
A similar generalization of monomer models has been
studied in the context of nonequilibrium wetting~\cite{Wetting}.
In both cases the additional evaporation process destroys the
stability of the smooth phase. In the dimer models, however, the roughening
transition of the dimer models is replaced by a faceting transition
between a rough and an (oppositely) faceted phase. Thus, the models of
Ref.~\cite{nohfac} link anomalous roughening and faceting transitions.

\section{Unidirectionally coupled branching-annihilating random walks 
	 with even number of offspring}
\label{CoupledSection}

\begin{figure}
\epsfxsize=85mm
\centerline{\epsffile{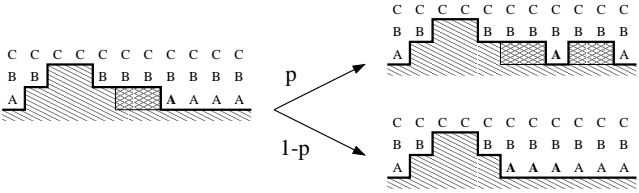}}
\vspace{2mm}
\caption{
\label{FigParticleInterpretation}
Extended particle interpretation. Dimers are adsorbed ($2A \rightarrow \O$)
and desorbed ($A \rightarrow 3A$) at the bottom layer. 
Similar processes take place at higher levels.
}
\end{figure}

In order to understand the universal properties of parity-conserving
roughening transitions let us extend the particle interpretation.
Assuming that $h=0$ is the bottom layer of the interface, we associate
particles $A,B,C,\ldots$ with sites at height $h_i \leq 0,1,2,\ldots$
(see Fig.~\ref{FigParticleInterpretation}).
As explained before, the dynamic processes at the bottom layer may be
interpreted as an effective reaction $A\rightarrow 3A, 2A\rightarrow \O$.
Similarly, the B-particles on top of the first layer
react by $B\rightarrow 3B, 2B\rightarrow \O$.
Clearly, the temporal evolution of the 
$B$-particles strongly depends on the actual configuration
of the $A$-particles. On the one hand, an $A$-particle
implies the presence of a $B$-particle at the same site,
giving rise to an effective reaction $A\rightarrow A+B$.
On the other hand, both the RSOS constraint and the restriction that
dimers react at adjacent sites of equal height may inhibit the
above reactions, introducing an effective feedback from higher levels
to lower ones. However, as suggested by our numerical results,
this type of inhibiting feedback does not affect
the critical behavior at the roughening transition.

Similarly, the $C$-particles perform an effective BAW2 on top of
the second layer. Therefore, the critical behavior of the first
few layers may be described in terms of a simplified
particle model where several BAW2 processes are {\em unidirectionally}
coupled according to the reaction scheme
\begin{eqnarray}
\label{ReactionScheme}
&&A \rightarrow 3A  \hspace{5.5mm} \qquad B \rightarrow 3B
\hspace{5.5mm}\qquad C \rightarrow 3C
\, \nonumber\\
&&2A \rightarrow \O \hspace{5.5mm} \qquad 2B \rightarrow \O
\hspace{5.5mm}\qquad 2C \rightarrow \O
\, \\
&&A \rightarrow A+B \qquad B \rightarrow B+C \qquad C \rightarrow C+D
\,, \ldots \nonumber
\end{eqnarray}
generalizing the concept of 'coupled DP'~\cite{CoupledDP}.
We propose this reaction scheme to characterize
the universal behavior of the dimer models at the
roughening transition.

\begin{figure}
\epsfxsize=85mm
\centerline{\epsffile{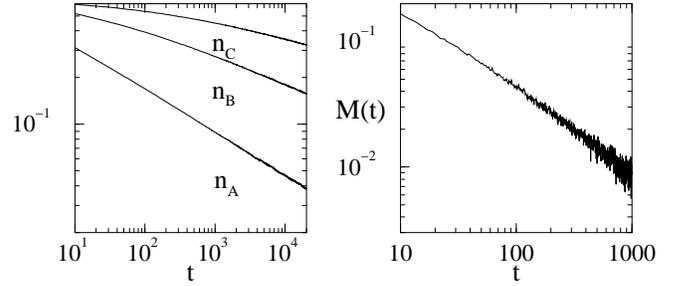}}
\vspace{2mm}
\caption{
Unidirectionally coupled PC processes. Left:
Densities $n_A,n_B,n_C$ as functions of time.
Right: Magnetization $M(t)$ as a function of time
averaged over $1000$ independent realizations.
\label{FigCoupled}
}
\end{figure}

To support this hypothesis, we study the
reactions~(\ref{ReactionScheme}) by Monte-Carlo simulations.
To this end we use three copies of the BAW2 introduced in Ref.~\cite{BAWMOD}
coupled by the rule that a selected $A$($B$)-particle instantaneously
creates a $B$($C$)-particle at the same position,
provided that the target site is empty. Using a lattice size of $2500$
sites we measure the particle densities $n_A,n_B,n_C$ as functions
of time at criticality.
As shown in Fig.~\ref{FigCoupled}, they display essentially
the same behavior as the densities $n_0,n_1,n_2$ in the dimer model.
Averaging over one decade in time we obtain the exponents
\begin{equation}
\delta_A = 0.280(5)
\ , \
\delta_B = 0.190(7)
\ , \
\delta_C = 0.120(10)
\ ,
\end{equation}
which are in fair agreement with the corresponding
exponents of the dimer model. We also measured
the analog of the magnetization $M=n_A-n_B+n_C-n_D+\ldots$
in $20$ coupled PC processes, obtaining an algebraic decay
$|M|\sim t^{-\theta/\nu_\parallel}$ 
with the exponent $\theta/\nu_\parallel =0.74(3)$.
This confirms the hypothesis that the roughening transition
of the dimer models belongs to the universality class of 
unidirectionally coupled PC processes. Notice that the
coupling $A\rightarrow A+B$ violates parity conservation
at the $B$-level. However, using a parity-conserving coupling
such as $A\rightarrow A+2B$ we obtain similar results at
the transition.

Before turning to the field-theoretic formulation of coupled
PC processes, let us consider the mean field approximation 
of the reaction scheme~(\ref{ReactionScheme})
\begin{eqnarray}
\partial_t n_A &=& \sigma n_A - \lambda n_A^2           \,, \nonumber \\
\partial_t n_B &=& \sigma n_B - \lambda n_B^2 + \mu n_A \,, \\
\partial_t n_C &=& \sigma n_C - \lambda n_C^2 + \mu n_B \,, \,\ldots\nonumber
\end{eqnarray}
where $n_A,n_B,n_C$ correspond to the
densities $n_0,n_1,n_2$ in the growth models.
$\sigma$ and $\lambda$ are  the rates for offspring production
(dimer evaporation) and pair annihilation (dimer deposition),
respectively. The coefficient $\mu$ is an effective coupling constant
between different particle species.
Since these equations are coupled in only one direction,
they can be solved by iteration. Obviously, the mean-field
critical point is $\sigma_c=0$. For small values of $\sigma$
the stationary particle densities in the active state are given by
\begin{equation}
n_A =      \frac{\sigma}{\lambda} \,, \,\,\,
n_B \simeq \frac{\mu}{\lambda} \left( \frac{\sigma}{\mu} \right)^{1/2} , \,\,\,
n_C \simeq \frac{\mu}{\lambda} \left( \frac{\sigma}{\mu} \right)^{1/4} \,,
\end{equation}
corresponding to the mean field critical exponents
\begin{equation}
\label{MeanFieldBeta}
\beta^{MF}_A=1 ,\,\,\,\,\beta^{MF}_B=1/2 ,\,\,\,\,\, \beta^{MF}_C=1/4
,\ldots \,.
\end{equation}
These exponents should be valid for $d>d_c=2$ (see below). 
Solving the asymptotic temporal behavior we
find $\nu_\parallel=1$, implying that $\delta^{MF}_k=2^{-k}$.
Although this simple mean-field calculation does
not include parity conservation, it explains
the reduced values of the exponents at higher levels.
The different numerical values in one dimension are due to
fluctuation corrections which may be computed within a field-theoretic
renormalization group approach. A field theory for a single BAW2
was introduced and studied in detail by Cardy and 
T\"auber~\cite{BAWEFT}. Following their notation, the effective
(unshifted) action of unidirectionally coupled BAW2's should be given by
\begin{eqnarray}
\label{EffectiveAction}
&&S[\psi_0,\psi_1,\psi_2,\ldots,\bar{\psi}_0,
\bar{\psi}_1,\bar{\psi}_2,\ldots]= \nonumber \\
&&\int d^dx\,dt\,\sum_{k=0}^\infty \Bigl\{
\bar{\psi}_k(\partial_t-D\nabla^2)\psi_k - \lambda(1-\bar{\psi}_k^2)\psi_k^2 +
\nonumber\\
&&\hspace{4mm}
+ \sigma(1-\bar{\psi}_k^2)\bar{\psi}_k\psi_k 
+ \mu (1-\bar{\psi}_k) \bar{\psi}_{k-1} \psi_{k-1}  \Bigr\}
\end{eqnarray}
where $\psi_{-1}=\bar{\psi}_{-1} \equiv 0$. Here 
the fields $\psi_k$ and $\bar{\psi_k}$ represent the
configurations of the system at level~$k$.
However, even for a single copy in 1+1 dimensions
the field-theoretic treatment poses considerable difficulties.
They stem from the presence of {\em two} critical dimensions:
$d_c=2$, above which mean-field theory applies,
and $d^\prime_c \approx 4/3$, where for $d>d^\prime_c$ ($d<d^\prime_c$)
the branching process is relevant (irrelevant) at the annihilation fixed
point. Therefore the physically interesting spatial dimension
$d=1$ cannot be accessed by a controlled $\epsilon$-expansion down
from upper critical dimension $d_c=2$. A one-loop calculation
for the action~(\ref{EffectiveAction}) would be an even more
difficult task.

A fundamental problem of the field theory for coupled DP
describing the monomer case is the relevant
coupling $\mu$ which {\em grows} under renormalization
group transformations~\cite{CoupledDP}. 
Moreover, even in a one-loop calculation 
certain infrared divergent diagrams are encountered 
which are proportional to $\mu$.
Similar difficulties are expected for unidirectionally coupled
PC processes. They might be responsible for violations of 
scaling in the long time limit appearing as a curvature
of $n_B$ and $n_C$ in Fig.~\ref{FigCoupled}.

\section{Parity-conserving polynuclear growth models}
\label{PNGSection}

Roughening transitions related to coupled DP were 
first observed in so-called polynuclear growth (PNG) 
processes~\cite{KW89,Richardson73,Goldenfeld84,KS89,LRWK90,Toom94}.
In these models the interface grows by nucleation processes
and deterministic growth of terraces. The use of parallel updates
ensures that the maximal propagation velocity is $1$. Depending on
the rates for nucleation and terrace growth, the models exhibit
a roughening transition from a moving rough phase to a smooth
phase propagating at maximal velocity. As shown in Ref.~\cite{Alon96},
PNG models may be viewed as upside-down versions of monomer
models in a co-moving frame. In some cases it is even possible to
relate PNG and monomer models exactly to each other, resolving the apparent
paradox that the transition in PNG models requires
parallel updates whereas for monomer models the
type of updates does not play a role.

\begin{figure}
\epsfxsize=75mm
\centerline{\epsffile{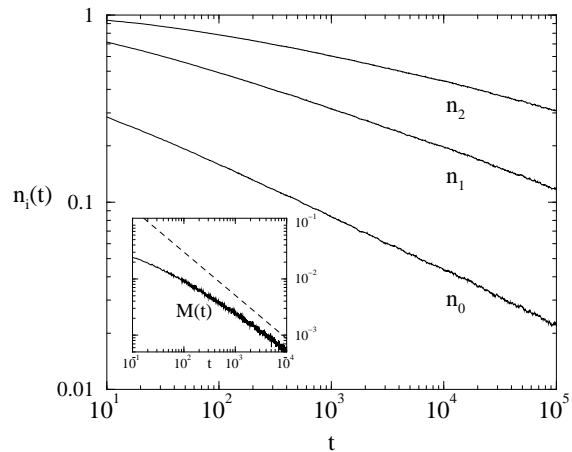}}
\caption{
\label{FigPng}
Parity-conserving PNG model.
The densities $n_0 \ldots n_3$ are shown as
functions of time. The exponents $\delta_k$
are estimated in the interval $10^3 \leq t \leq 10^5$,
suggesting that the model belongs to the universality class
of unidirectionally coupled PC processes. The inset shows
the magnetization $M(t)$. The dashed line indicates the
slope $-0.77$.
}
\end{figure}

In order to understand the DP mechanism,
let us consider the PNG model introduced by
Kert\'esz and Wolf~\cite{KW89}
which is updated synchronously in two
sub-steps. At first all up (down) steps of the interface move
deterministically to the left (right) over a distance of $u$
lattice spacings. Then all heights are increased with
probability~$p$. Thus the model is unrestricted. For
$u=1$ this dynamic rule can be expressed as a single parallel
update
\begin{equation}
h_i(t+1) =
\left\{
\begin{array}{ll}
m(t)+1 & \mbox{ with prob. $p$ } \\
m(t)   & \mbox{ with prob. $1-p$ }
\end{array}
\right. \, ,
\end{equation}
where $m(t)=\max[h_{i-1}(t),h_{i}(t),h_{i+1}(t)]$.
Starting from a flat interface $h_i(0)=0$, the sites at
maximal height $h_i(t)=t$ may be considered as the active
sites of a DP process. The nucleation process
turns active into inactive sites with probability $1-p$
while the deterministic growth of terraces resembles
offspring production. Therefore, if $p$ is large enough, the
interface is smooth and propagates with velocity $1$. 
Below a critical threshold $p_c=0.539(1)$, however,
the growth velocity is smaller than $1$ and the interface
evolves into a rough state.
As shown in Ref.~\cite{CoupledDP}, this model
is a realization of unidirectionally coupled DP
processes in a co-moving frame,
where the order parameters $n_k$ are given by
\begin{equation}
\label{PNGObservable}
n_k(t) = \frac{1}{L} \sum_i \sum_{h=0}^{k} \delta_{h_i(t),t-h} \ .
\end{equation}
In the following we introduce a
parity-conserving PNG model which belongs to the universality
class of unidirectionally coupled PC processes. The model is defined 
on a one-dimensional lattice with periodic boundary conditions and
evolves by sublattice-parallel updates. In the first half 
time step pairs of sites $(i,i+1)$ with even $i$ are updated.
If $h_i(t) \neq h_{i+1}(t)$, the heights are incremented by one step
\begin{eqnarray}
\label{incrule}
h_i(t+1/2)&=&h_i(t)+1 \ , \\
h_{i+1}(t+1/2)&=&h_{i+1}(t)+1 \ .\nonumber
\end{eqnarray}
If, however, the two heights are equal, they are
updated by the probabilistic rule
\begin{eqnarray}
h_i(t+1/2) &=& h_{i+1}(t+1/2)= \nonumber \\
&=&
\left\{
\begin{array}{ll}
m(t)+1 & \mbox{ with prob. $p$ } \\
m(t)   & \mbox{ with prob. $1-p$ } 
\end{array}
\right. \, ,
\end{eqnarray}
where $m(t)=\max[h_{i-1}(t),h_{i}(t),h_{i+1}(t),h_{i+2}(t)]$.
In the second half time step the same update rule is applied to
odd pairs of sites. Clearly, this model generalizes the PNG model
of Ref.~\cite{KW89} and conserves parity at each height level
in a co-moving frame. The conservation law leads to the formation
of pinning centers moving at maximal velocity. 

Performing Monte-Carlo simulations we observe a roughening transition at 
the critical threshold $p_c=0.5697(3)$. Starting from a flat interface,
we measure the densities $n_k(t)$ defined in
Eq.~(\ref{PNGObservable}) as functions of time.
As shown in Fig.~\ref{FigPng}, the temporal decay of $n_k$ is 
similar to the one observed in the dimer model. 
Averaged over two decades in time we obtain the estimates
\begin{equation}
\delta_0=0.28(1), \,\,\,\,
\delta_1=0.21(2), \,\,\,\,
\delta_2=0.14(2), 
\end{equation}
which are compatible with the values listed in Table~\ref{Table}.
The magnetization $M(t)$ defined in Eq.~(\ref{Magnetization}) 
does not show a clean power law behavior but seems to approach
an asymptotic decay $t^{-0.77}$ in agreement with previous findings
(see inset of Fig.~\ref{FigPng}). Therefore, we conclude that
the roughening transition of the parity-conserving PNG model 
belongs to the universality class of unidirectionally 
coupled PC processes.

\section{Conclusions and outlook}
\label{ConcSection}

In this paper we have investigated a class of parity-conserving growth
processes in which dimers adsorb at sites of equal height and
desorb at the edges of terraces. At a critical growth
rate $p=p_c$ the models display a roughening transition from a
smooth to a rough phase. In order to demonstrate the robustness
of this transition, we have studied four variants of a 1+1-dimensional
parity-conserving growth process with and without RSOS constraint,
using either random-sequential or sublattice-parallel dynamics.
In addition, we have introduced a parity-conserving polynuclear growth
process where a similar transition takes place in a co-moving frame.
In all cases the roughening transition is characterized by the same
type of critical behavior.

The investigated dimer models generalize previously studied
monomer models. Their essential feature is a parity conservation
law at each height level, changing the universal properties
of the roughening transition. The conservation law leads to the
formation of pinning centers separating regions of even and
odd parity. Thus, in contrast to the monomer case, the interface
remains pinned to the initial height level.
At the transition the width is found to increase {\em logarithmically}
with time due to a slow diffusion and annihilation of the pinning centers.
In Eq.~(\ref{WidthFSScaling}) we have proposed a finite-size scaling
form for logarithmic roughening which is confirmed by high precision
simulations. Moreover, we have shown that the
universal critical behavior at the first few layers may be
described in terms of unidirectionally coupled branching-annihilating
random walks with two offspring. Thus we suggest that the concept
of undirectionally coupled PC processes defines a whole universality
class of parity-conserving roughening transitions.

In case of equal rates for dimer adsorption and desorption
all variants undergo a second phase transition where the
width increases {\em algebraically}. In the restricted case
this transition has been identified as a faceting transition
from a rough to a faceted phase~\cite{nohfac}.
In the unrestricted variants, however, we observe a
transition from a rough to a freely growing phase characterized
by spiky interface configurations.

There are various possible extensions and generalizations of the
models studied in this paper. Very recently, Noh {\it et al.}~\cite{nohfac}
investigated a generalization of variant A where dimers may also
evaporate from the middle of plateaus. Even at a very small rate
this additional process destroys the stability of the smooth phase,
turning it into a faceted phase. Remarkably, a sharp transition
between the faceted and the rough phase still remains,
leading to interesting crossover phenomena between different
universality classes which have not been studied so far.
It would also be interesting to investigate parity-conserving
growth processes in higher dimensions. Since the upper critical
dimension $d^\prime_c$ is less than 2, we expect the roughening transition
-- if still existing -- to be described by mean-field exponents.
One may also consider growth processes of $n$-mers where the number
of particles at each height level is preserved modulo $n$.
Especially in higher dimensions, these $n$-mers might appear in
different shapes and orientations. After all it would also be
interesting to find experimental realizations for deposition and
evaporation of composite particles.

\vspace{3mm}
\noindent
Acknowledgements:\\
We would like to thank N. Menyh\'ard for pointing out the possibility
of parity-conserving polynuclear growth processes.
H.H. thanks the MTA-MFA in Budapest for hospitality where
parts of the work have been done.
The simulations were performed partially on the FUJITSU AP-1000,
AP-3000 and System-V parallel supercomputers. 
We thank R. Bishop for helping us to
operate the System-V machine. G.\'O. gratefully
acknowledges support from the Hungarian research fund OTKA
(Nos. T025286 and T023552).



\begin{thebibliography}{99}
\bibitem{Comment1}
	In \cite{HinrichsenOdor99a} we introduced this scaling form
	as $W^2(L,t) \simeq \ln [t^a f(t/L^z)]$.
	Here we use a slightly different notation in order to emphasize
	that $a$ is an amplitude rather than a critical exponent.
\bibitem{Comment2}
	Our earlier conjecture $W \sim \ln t$ in Ref.
	\cite{HinrichsenOdor99a} could not be
	confirmed by more precise simulations.
\end{thebibliography}
\end{document}